\documentclass[aip,prb,twocolumn,showpacs]{revtex4}

\usepackage{graphics,graphicx,amssymb,amsmath,color}
\begin{document}

\title{Evolution of the commensurate and incommensurate magnetic phases of the $S$~=~3/2 kagome staircase Co$_{3}$V$_{2}$O$_{8}$ in an applied field}
\author{Joel~S.~Helton$^{1,*}$}
\author{Ying~Chen$^{1,2}$}
\author{Georgii~L.~Bychkov$^{3}$}
\author{Sergei~N.~Barilo$^{3}$}
\author{Nyrissa~Rogado$^{4,\dag}$}
\author{Robert~J.~Cava$^{4}$}
\author{Jeffrey~W. Lynn$^{1,\ddag}$}

\affiliation{$^{1}$NIST Center for Neutron Research, National Institute of Standards and Technology, Gaithersburg, MD 20899, USA}
\affiliation{$^{2}$Department~of~Materials~Science~and~Engineering,~University~of~Maryland,~College~Park,~MD~20742,~USA}
\affiliation{$^{3}$Institute of Solid State and Semiconductor Physics, Academy of Sciences, Minsk 220072, Belarus}
\affiliation{$^{4}$Department of Chemistry, Princeton University, Princeton, NJ 08544, USA}
\date{\today}
\begin{abstract}
Single crystal neutron diffraction studies have been performed on the $S$~=~3/2 kagome staircase compound Co$_{3}$V$_{2}$O$_{8}$ with a magnetic field applied along the magnetization easy-axis ($\vec{H}$~$||$~$\vec{a}$).  Previous zero-field measurements [Y. Chen, \emph{et al.} Phys. Rev. B {\bf74}, 014430 (2006)] reported a rich variety of magnetic phases, with a ferromagnetic ground state as well as incommensurate, transversely polarized spin density wave (SDW) phases [with a propagation vector of $\vec{k}$~=~(0~$\delta$~0)] interspersed with multiple commensurate lock-in transitions.  The magnetic phase diagram with $\vec{H}$~$||$~$\vec{a}$ adds further complexity.  For small applied fields, $\mu_{0}H$~$\approx$~0.05~T, the commensurate lock-in phases are destabilized in favor of the incommensurate SDW ones, while slightly larger applied fields restore the commensurate lock-in phase with $\delta$~=~1/2 and yield a new commensurate phase with $\delta$~=~2/5.  For measurements in an applied field, higher-order scattering is observed that corresponds to the second-harmonic.

\end{abstract}
\maketitle

\section{INTRODUCTION}

\begin{figure}
\centering
\includegraphics[width=8.3cm]{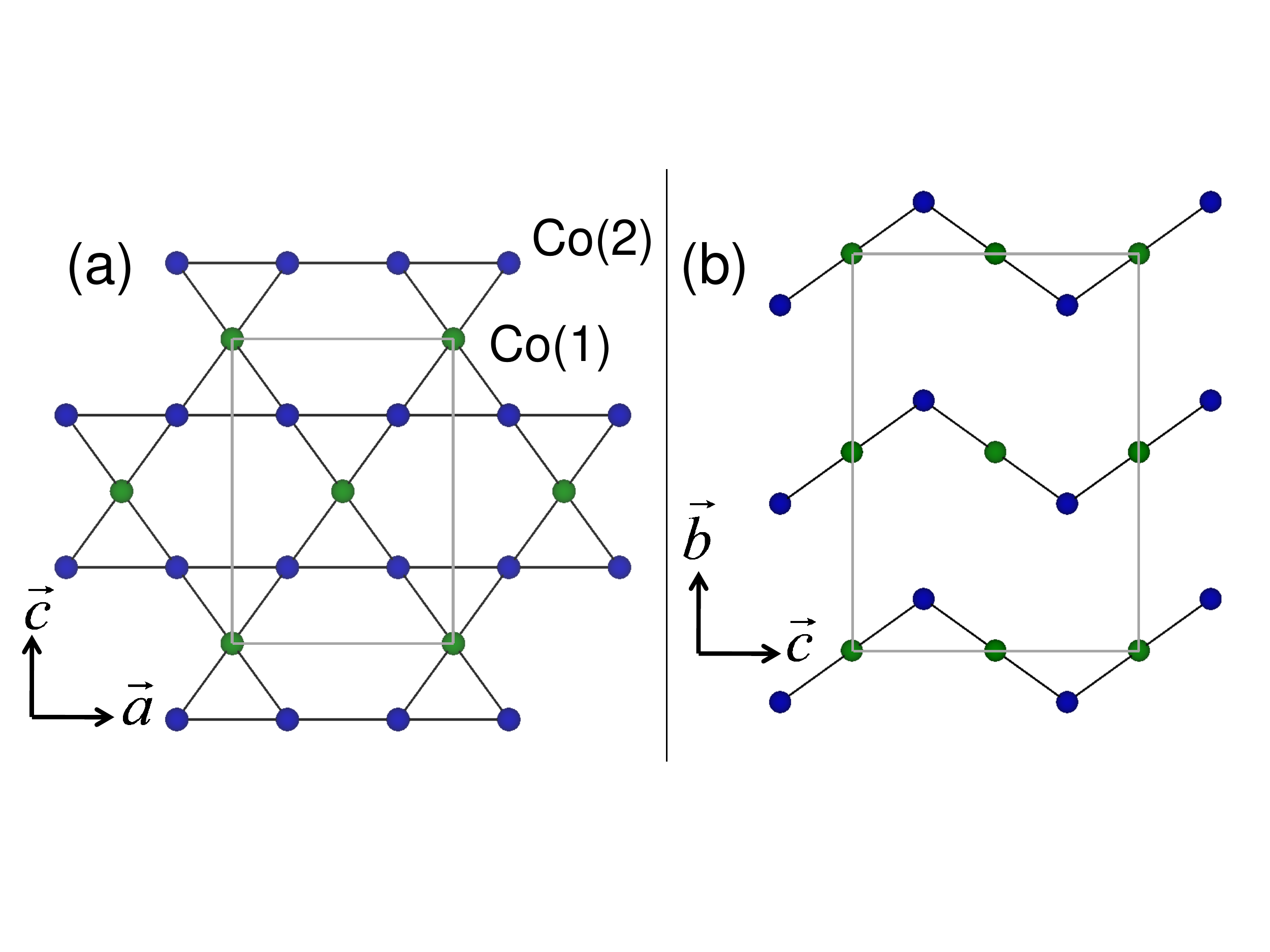} \vspace{-13mm}
\caption{The kagome staircase structure of Co$_{3}$V$_{2}$O$_{8}$.  Co$^{2+}$ ions form buckled kagome sheets in the $a$-$c$ plane that are stacked along the $b$-axis.  The Co(2) ``spine'' sites (shown in blue) form chains along the $a$-axis that are coupled through the Co(1) ``cross-tie'' sites (green). (a) Kagome structure in the $a$-$c$ plane.  (b) Buckled planes stacked along the $b$-axis.  The unit cell is outlined in gray.} \vspace{-2mm}
\label{Lattice}
\end{figure}

Frustrated magnets - systems where lattice geometry\cite{Ramirez}, disorder\cite{Villain}, or competing interactions\cite{Yoshimori} prevent a simple magnetic structure that satisfies all exchange interactions - often develop exotic quantum ground states and low-energy excitations.  One geometrically frustrated lattice of considerable interest is the kagome lattice\cite{Syozi,Elser}, consisting of corner sharing equilateral triangles.  A variation on this structure, the kagome staircase lattice, has been realized in the $M_{3}$V$_{2}$O$_{8}$ compounds where the divalent cation $M$~=~Mn, Co, Ni, Cu, or Zn\cite{Rogado2002,Rogado2003,Balakrishnan,Lawes,Szymczak,Morosan}.  These materials display buckled kagome planes stacked along the $b$-axis.  As shown in Figure~\ref{Lattice}, this structure results in two crystallographically distinct $M$ sites: the $M$(1) sites (at the 4$a$ Wyckoff positions) are referred to as the ``cross-tie'' sites while the $M$(2) sites (8$e$ positions) are the ``spine'' sites.  Chains of spine sites run along the $a$-axis, and are coupled through the cross-tie sites.  The complicated interplay of competing magnetic interactions in Ni$_{3}$V$_{2}$O$_{8}$ (NVO) yields a rich magnetic phase diagram\cite{Lawes,Kenzelmann}, with several distinct commensurate and incommensurate spin structures arising in the order of the $S$~=~1 Ni$^{2+}$ moments.  NVO is multiferroic\cite{Cabrera}, as the low-temperature incommensurate magnetic structure in this material is a spiral structure that breaks spatial inversion symmetry and leads to simultaneous ferroelectric order\cite{Lawes2005,Harris}.  Co$_{3}$V$_{2}$O$_{8}$ (CVO) features the kagome staircase lattice decorated with $S$~=~3/2 Co$^{2+}$ ions.  The crystal structure is orthorhombic, with $Cmca$ space group and lattice parameters of $a$~=~6.027~{\AA}, $b$~=~11.483~{\AA}, and $c$~=~8.296~{\AA}.  Much of the interest in kagome lattice materials\cite{Broholm,Hiroi,Helton,Mendels} stems from theoretical and numerical predictions\cite{Zeng,Singh,Sachdev} of a disordered quantum ground state in the low-spin kagome lattice Heisenberg antiferromagnet.  In the kagome staircase compounds Ni$_{3}$V$_{2}$O$_{8}$ and Co$_{3}$V$_{2}$O$_{8}$, the crystallographically distinct spine and cross-tie sites interact through a variety of exchange interactions both ferromagnetic and antiferromagnetic, including interactions beyond the nearest-neighbor.  Therefore NVO and CVO are of interest in exploring the effects of multiple competing magnetic interactions in a frustrated geometry.

Previous single crystal neutron diffraction experiments\cite{Chen} on CVO in zero field revealed a rich variety of magnetic phases.  Below $T_{C}$~=~6.2~K the system is ferromagnetically ordered, with easy magnetization along the $a$-axis.  A magnetic field of $\mu_{0}H$~=~0.3~T or greater applied along the $a$-axis in this low-temperature ferromagnetic phase will produce a saturated magnetization\cite{Szymczak,WilsonJPCM} of about 3~$\mu_{B}$/Co; recent measurements suggest that a portion of this moment resides on the O or V sites\cite{Qureshi2009}.  Above the ferromagnetic ground state, magnetic reflections were observed at (0~$n \pm \delta$~$l$) where $n$ is an even integer, $l$ is an integer, and $\delta$ evolves from 1/3 just above the ferromagnetic phase to 0.55 at $T_{N}$~=11.3~K, above which the system is paramagnetic.  These reflections reveal a transversely polarized antiferromagnetic (AF) spin density wave (SDW) state with a propagation vector of $\vec{k}$~=~(0~$\delta$~0).  The value of $\delta$ displays commensurate lock-in phases at 1/3 ($T_{C}$~$<$~$T$~$<$~6.5~K) and 1/2 (6.9~K~$<$~$T$~$<$~8.6~K) interspersed with incommensurate SDW phases of 1/3~$<$~$\delta$~$<$~1/2 and 1/2~$<$~$\delta$~$\leq$~0.55.  In the ferromagnetic ground state, the ordered moments on the spine and cross-tie sites were refined as 2.73~$\mu_{B}$ and 1.54~$\mu_{B}$, respectively.  The magnetic structure in the $\delta$~=~1/2 AF phase consists of alternating ferromagnetic and antiferromagnetic layers; in the ferromagnetic layers the spine and cross-tie sites have ordered moments of 1.39~$\mu_{B}$ and 1.17~$\mu_{B}$, respectively, while in the antiferromagnetic layers the spine sites display an ordered moment of 2.55~$\mu_{B}$ and the cross-tie sites are fully frustrated with no ordered moment.  The spin direction for all Co moments is along the crystallographic $a$-axis (the magnetization easy-axis), showing strong Ising-like behavior.  Magnetic phase diagrams for CVO with $\vec{H}$ parallel to the $b$- and $c$-axes have been reported\cite{Yasui,WilsonJPCM,Qureshi2007}, with a complex evolution of the magnetic phases in field.

The crystallographic structures of Co$_{3}$V$_{2}$O$_{8}$ and Ni$_{3}$V$_{2}$O$_{8}$ are remarkably similar, yet the magnetic structures displayed by the two materials are quite different.  These differences arise from the fact that the magnetic structures in CVO and NVO are determined by the subtle competition between  nearly balanced countervailing magnetic interactions in a frustrated geometry.  As the structure consists of edge-sharing $M$O$_{6}$ octahedra, the $M$-O-$M$ bond angles are close to 90$^{\circ}$ and nearest-neighbor superexchange interactions will be fairly weak so that interactions beyond the nearest-neighbor play a significant role.  The zero-field magnetic structures of Co$_{3}$V$_{2}$O$_{8}$ have been qualitatively reproduced by a minimal interaction Ising model with four competing spine-spine interactions\cite{Chen}, including a temperature dependent interaction between nearest-neighbor spines that is influenced by the cross-tie Co(1) moments.  While recent spin wave measurements\cite{Ramazanoglu} in the low-temperature ferromagnetic phase were modeled with only a ferromagnetic nearest-neighbor Co(1)-Co(2) interaction, the strengths of the competing spine-spine interactions that determine the antiferromagnetic structures are not yet known.

\begin{figure}
\vspace{2mm}
\centering
\includegraphics[width=8.8cm]{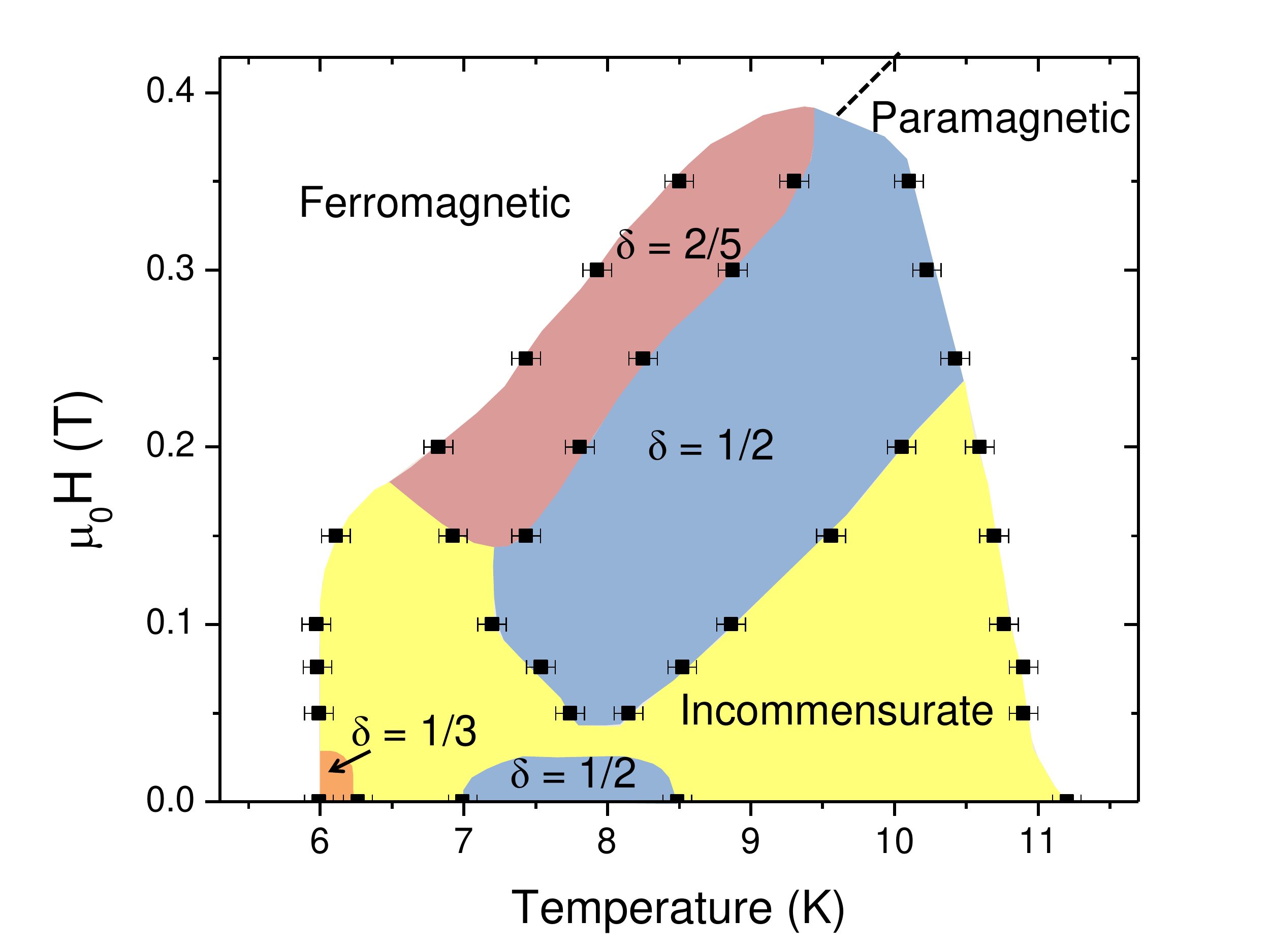} \vspace{-3mm}
\caption{$H$-$T$ phase diagram for Co$_{3}$V$_{2}$O$_{8}$ with $\vec{H}$~$||$~$\vec{a}$.  The material is paramagnetic at high temperature, and ferromagnetic at low temperature or under a large applied field.  The antiferromagnetic phases are described by a transversely polarized spin density wave with moments aligned parallel to $\vec{a}$ and a propagation vector along $\vec{b}$.  Incommensurate SDW phases are displayed, as well as commensurate lock-in phases at $\delta$~=~1/3, 2/5 and 1/2.} \vspace{-6mm}
\label{PhaseDiagram}
\end{figure}

In this paper, we present single crystal neutron diffraction data on Co$_{3}$V$_{2}$O$_{8}$ measured in small magnetic fields with $\vec{H}$~$||$~$\vec{a}$.  We have determined the $H$-$T$ magnetic phase diagram, as is shown in Figure~\ref{PhaseDiagram}.  We find that a weak applied field, $\mu_{0}H$~$\approx$~0.05~T, destabilizes the commensurate lock-in phases in favor of the incommensurate AF phase.  Slightly higher magnetic fields restore the stability of the $\delta$~=~1/2 AF phase, and for applied fields greater than 0.15~T we find a new commensurate antiferromagnetic phase with $\delta$~=~2/5.  No antiferromagnetic phases are observed for $\mu_{0}H$~$\geq$~0.4~T; the high-field transition between the ferromagnetic and paramagnetic phases shown in Figure~\ref{PhaseDiagram} is estimated from previously reported data\cite{Yasui}.  The $H$-$T$ phase diagram of CVO with $\vec{H}$~$||$~$\vec{a}$ has been previously reported\cite{Qureshi2007,Yasui,Yen} on the basis of bulk magnetization, specific heat, and dielectric measurements.  Our results are in general agreement with these reports on the extent of the ferromagnetic and paramagnetic phases; however, the use of single crystal neutron diffraction has enabled us to better elucidate the evolving propagation vector and discern the new AF phase with $\delta$~=~2/5.  For measurements in an applied field of at least 0.05~T we find second-order scattering at (0~$n \pm \delta^{\prime}$~$l$) positions where $\delta^{\prime}$~=~2$\delta$. This is a clear change from the third-order scattering observed in zero field\cite{Chen}, and is possible only in the presence of a net magnetization.

\section{EXPERIMENT}

Neutron diffraction experiments on single crystal samples of Co$_{3}$V$_{2}$O$_{8}$ were performed on the BT-7 and BT-9 thermal triple-axis neutron spectrometers at the NIST Center for Neutron Research.  The growth of single crystal samples has been reported previously\cite{Chen}.  Samples were mounted in the (0~$K$~$L$) scattering plane and placed inside a helium flow dewar inserted into a 7~T vertical-field superconducting magnet.  Thus the magnetic field was applied parallel to the crystallographic $a$-axis, which is also the magnetization easy-axis. Measurements were performed on both a small ($\approx$3$\times$3$\times$2~mm$^{3}$) and a large (4.6~g, roughly 1~cm on a side) single crystal.  The small crystal was measured on BT-7 while the large crystal was measured on BT-9.  The (0~0~2) reflections of the pyrolytic graphite (PG) monochromator and analyzer selected the fixed neutron energy of 14.7~meV ($\lambda$~=~2.36~{\AA}) and PG filters were used to reduce higher-order neutrons.  When a field is applied parallel to the $a$-axis in the low-temperature ferromagnetic phase we find a decrease in the intensity of ferromagnetic Bragg reflections, with the magnetic scattering intensity strongly suppressed under applied fields of 0.1~T or greater.  This behavior has been previously reported by Qureshi, \emph{et al.}\cite{Qureshi2009} and like that report we attribute this effect to increasing primary extinction as the size of individual magnetic domains grows upon application of a field.

There were quantitative differences in the scattering observed from the small and large crystals, with a contrast in the boundaries between incommensurate phases and commensurate lock-ins.  The zero-field $\delta$~=~1/3 phase, present over only a 0.3~K~temperature range in the small crystal, was not observed as a clear lock-in phase in the large crystal and at low fields the $\delta$~=~1/2 phase was observed over a narrower temperature range and at a slightly elevated temperature in the large crystal when compared to the small crystal.  Qualitatively, measurements on both crystals were otherwise in good agreement, displaying the same magnetic phases and higher-order scattering.  The $H$-$T$ phase diagram (shown in Figure~\ref{PhaseDiagram}) was determined using data from the small crystal, which is likely to be of higher quality; this crystal was also used for previously reported zero-field single crystal diffraction measurements\cite{Chen}.  The data displayed in most other figures represent measurements on the large crystal, as the increased scattering intensity allows us to more clearly observe the higher-order peaks; however, the field dependence of the (0~3~0) reflection at 7.9~K was measured on the small crystal.

\section{RESULTS}

\subsection{Magnetic Phase Diagram}

\begin{figure}
\centering
\vspace{2mm}
\includegraphics[width=8.3cm]{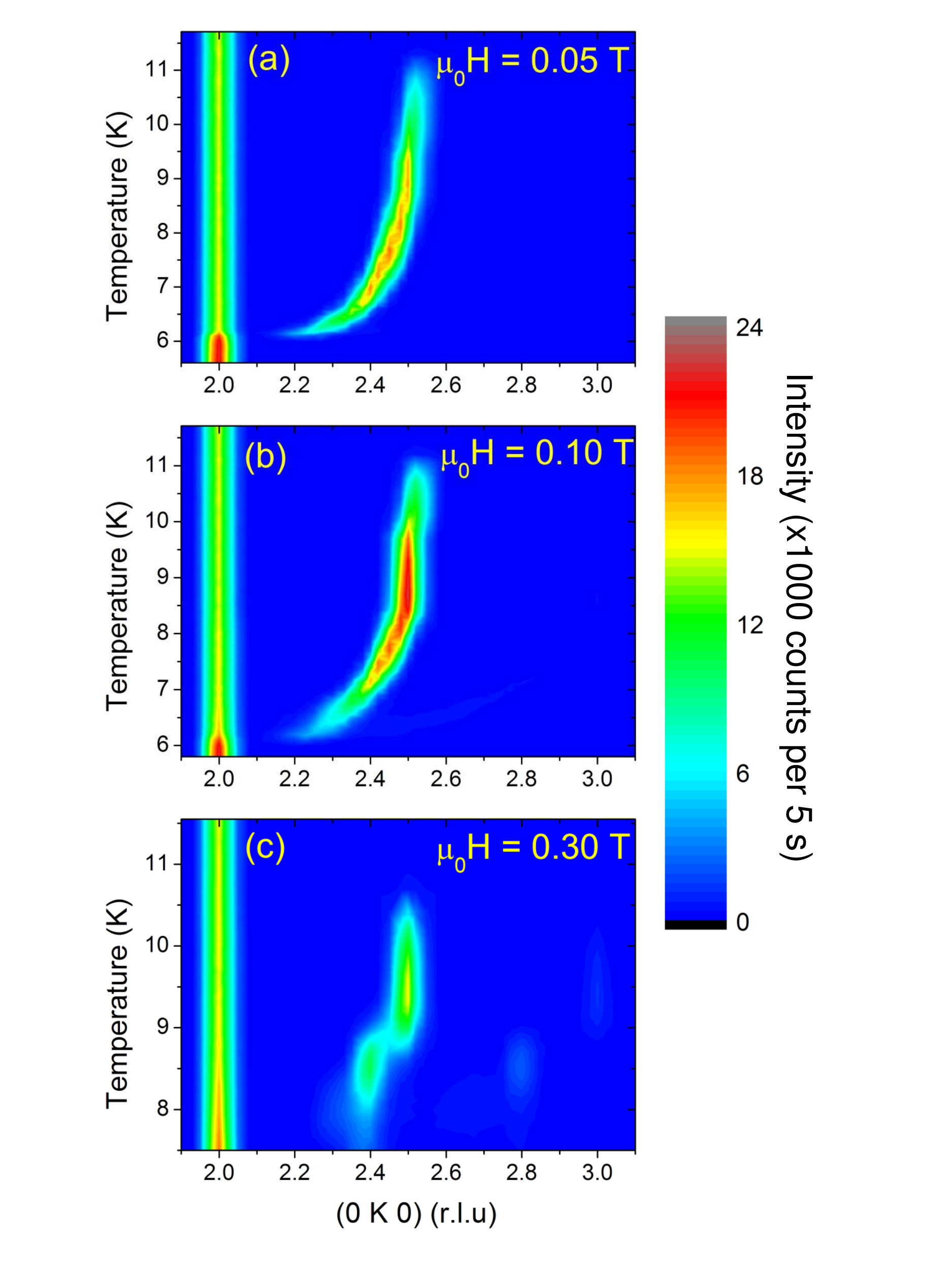} \vspace{-4mm}
\caption{Intensity color plots for scattering under applied fields of 0.05~T to 0.30~T.  (a) Scattering with $\mu_{0}H$~=~0.05~T.  Incommensurate phases dominate this scattering, with the $\delta$~=~1/2 AF phase stable over only a narrow temperature range. (b) Scattering with $\mu_{0}H$~=~0.10~T.  The $\delta$~=~1/2 commensurate AF phase is again stable over a temperature range of about 2~K.  (c) Scattering with $\mu_{0}H$~=~0.30~T.  The data display only two commensurate phases, with $\delta$~=~2/5 and $\delta$~=~1/2.} \vspace{-2mm}
\label{Mesh}
\end{figure}

Diffraction measurements were performed at several magnetic field values between $\mu_{0}H$~=~0.05~T and 0.40~T.  For each field, diffraction scans were measured with a temperature increase between successive scans of 0.1~K after first field-cooling from the paramagnetic phase into the ferromagnetic phase.  All measurements on CVO reveal a propagation vector parallel to $\vec{b}^{\ast}$ such that only scans in the (0~$K$~0) direction will be considered.  A few representative measurements are shown in Figure~\ref{Mesh}; data are shown for $\mu_{0}H$~=~0.05~T~(panel~(a)), 0.10~T~(panel~(b)), and 0.30~T (panel~(c)).  The data in Figure~\ref{Mesh}(a), with $\mu_{0}H$~=~0.05~T, display significant differences from the zero-field behavior. This small magnetic field has destabilized the commensurate lock-in phases, as the $\delta$~=~1/2 lock-in is present over only a very narrow temperature range.  Similar to the zero-field case, an incommensurate propagation vector is observed for temperatures above and below the commensurate lock-in phase.  The incommensurate scattering at temperatures below the $\delta$~=~1/2 lock-in evolves with a propagation vector 0.27~$\leq$~$\delta$~$<$~1/2; this stands in contrast with the zero-field case where the smallest $\delta$ value measured was in the $\delta$~=~1/3 lock-in phase.  At temperatures above the $\delta$~=~1/2 lock-in phase the incommensurate propagation vector takes values of 1/2~$<$~$\delta$~$\leq$~0.55, similar to the zero-field data.

With an applied field of $\mu_{0}H$~=~0.10~T, shown in Figure~\ref{Mesh}(b), the $\delta$~=~1/2 phase is stable over a temperature range comparable to that observed in zero field.  This behavior is consistent with recently reported data from Petrenko, \emph{et al.}, where measurements at 7.5~K displayed a commensurate $\delta$~=~1/2 propagation vector at zero field and at $\mu_{0}H$~=~0.15~T, yet displayed an incommensurate $\delta$~=~0.475 propagation vector at $\mu_{0}H$~=~0.072~T\cite{Petrenko}.  It is possible that the low-field incommensurate structure measured here arises from long wavelength ``spin slips"\cite{Gibbs} similar to the behavior reported for ErNi$_{2}$B$_{2}$C\cite{Choi,Lynn,Jensen}.  An intriguing question remains whether the two regions of the $H$-$T$ phase diagram displaying a $\delta$~=~1/2 propagation vector are connected or distinct.  The phase diagram with $\vec{H}$~$||$~$\vec{a}$ determined by Yen, \emph{et al.}\cite{Yen} reported dielectric constant anomalies at small applied fields, signalling the loss of the zero-field $\delta$~=~1/2 phase for all temperatures where this phase was present.  Further, in field dependent neutron diffraction measurements the intensity of the (0~3~0) reflection, which will be shown to be the higher-order reflection of the high-field $\delta$~=~1/2 phase, at 7.9~K displays a sharp transition at $\mu_{0}H$~=~0.05~T suggesting that these phases are distinct.

Data measured with $\mu_{0}H$~=~0.30~T, as shown in Figure~\ref{Mesh}(c), display only two phases: the $\delta$~=~1/2 phase as well as a new commensurate lock-in phase with $\delta$~=~2/5.  This new phase corresponds to a spin density wave with a wavelength of five kagome layers, intermediate to the two previously known commensurate lock-in phases which have wavelengths of four ($\delta$~=~1/2) or six ($\delta$~=~1/3) layers.  Extensive neutron diffraction measurements\cite{Yasui,Petrenko} have been reported with $\vec{H}$~$||$~$\vec{b}$ and $\vec{H}$~$||$~$\vec{c}$; a commensurate lock-in with $\delta$~=~2/5 has not been reported in any of these measurements and appears to be unique to fields along the magnetization easy-axis.  Over the full range of fields for which this  $\delta$~=~2/5 phase is present (0.15~T~$\leq$~$\mu_{0}H$~$\leq$~0.35~T), it is bounded on the high-temperature side by the $\delta$~=~1/2 phase; this phase transition does not feature any intervening incommensurate phase with a $\delta$ value between 2/5 and 1/2.  In zero field, the incommensurate propagation vector passes through a value of $\delta$~=~2/5 with no sign of a commensurate lock-in; this suggests that the $\delta$~=~2/5 lock-in phase in field is stabilized not by higher-order terms in the Landau expansion of the Hamiltonian, as is the case for the zero-field lock-ins\cite{Chen}, but rather by the Zeeman energy.  It should be noted that neither an incommensurate square density wave nor a commensurate square density wave where the wavelength is an even number of kagome layers could yield a net moment in this system.  However, a commensurate spin density wave whose wavelength is an odd number of kagome layers can develop a net moment through a squaring-up of the sinusoidal SDW; in particular, a SDW with $\delta$~=~2/5 can evolve towards an up-up-up-down-down state with a net moment.

\subsection{Second-Order Peaks}
\begin{figure}
\centering
\vspace{2mm}
\includegraphics[width=8.0cm]{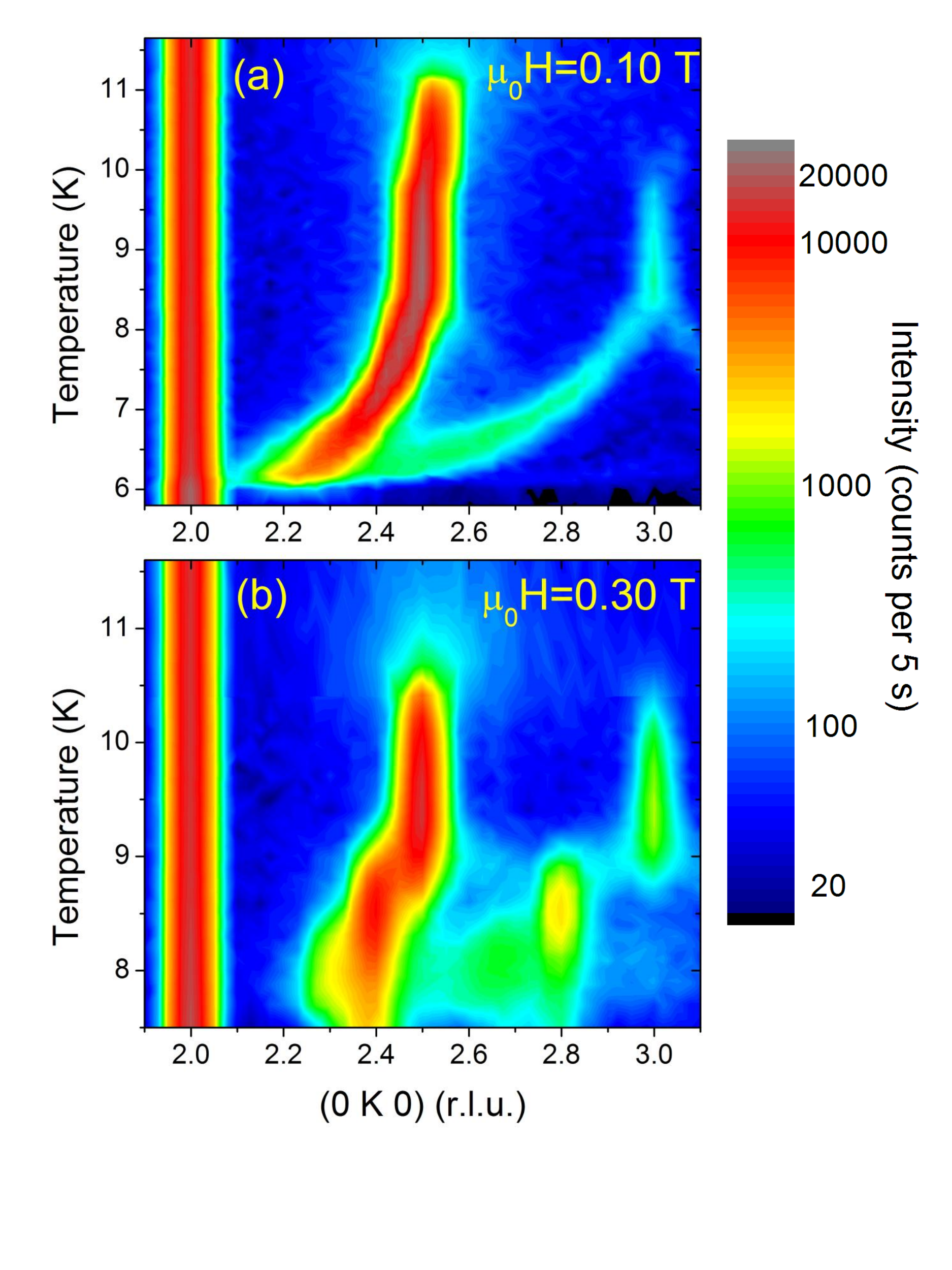} \vspace{-12mm}
\caption{Intensity color plots with a logarithmic scale for scattering at fields $\mu_{0}H$~=~0.10~T (panel a) and $\mu_{0}H$~=~0.30~T (panel b).  The logarithmic scale is used to highlight the second-order scattering at (0~2+$\delta^{\prime}$~0).} \vspace{-2mm}
\label{Mesh_HO}
\end{figure}
Zero-field measurements on CVO\cite{Chen} found, upon cooling, incommensurate reflections with 1/3~$<$~$\delta$~$<$~1/2 for 6.5~K~$<$~$T$~$<$~6.9~K and the $\delta$~=~1/3 phase for 6.2~K~$<$~$T$~$<$~6.5~K.  These phases also displayed higher-order scattering in addition to the first-order reflection at (0~$\delta$~0), with a second peak observed at (0~$\delta^{\dag}$~0) where $\delta^{\dag}$~=~2-3$\delta$.  Alternatively, one can express these positions by defining the fundamental peak position as $\delta$~=~1/2-$\xi$; the higher-order peak will then be located at $\delta^{\dag}$~=~1/2+3$\xi$.  This relation persists into the $\delta$~=~1/3 phase where a strong higher-order peak was observed at (0~1~0).  Higher-order scattering is common when a spin density wave deviates from a simple sinusoid by ``squaring up''; even-order harmonics are forbidden when the system does not feature a net magnetic moment, which is consistent with the third-order peak observed in the zero-field data.  Figure~\ref{Mesh_HO} displays the data measured under applied fields of 0.10~T (panel~(a)) and 0.30~T (panel~(b)) with a logarithmic intensity scale.  With the logarithmic scale, higher-order scattering is observed at (0~2+$\delta^{\prime}$~0) in addition to the primary peak at (0~2+$\delta$~0).  For $\mu_{0}H$~=~0.10~T the higher-order peak emerges from the primary peak at low temperatures, and $\delta^{\prime}$ increases with increasing temperature until $\delta^{\prime}$~=~1 in the phase where the primary peak position is described by $\delta$~=~1/2.  For $\mu_{0}H$~=~0.30~T the higher-order peak is located at (0~2.8~0) or (0~3~0) in, respectively, the phases where $\delta$~=~2/5 or 1/2.

\begin{figure}
\centering
\vspace{2mm}
\includegraphics[width=7.0cm]{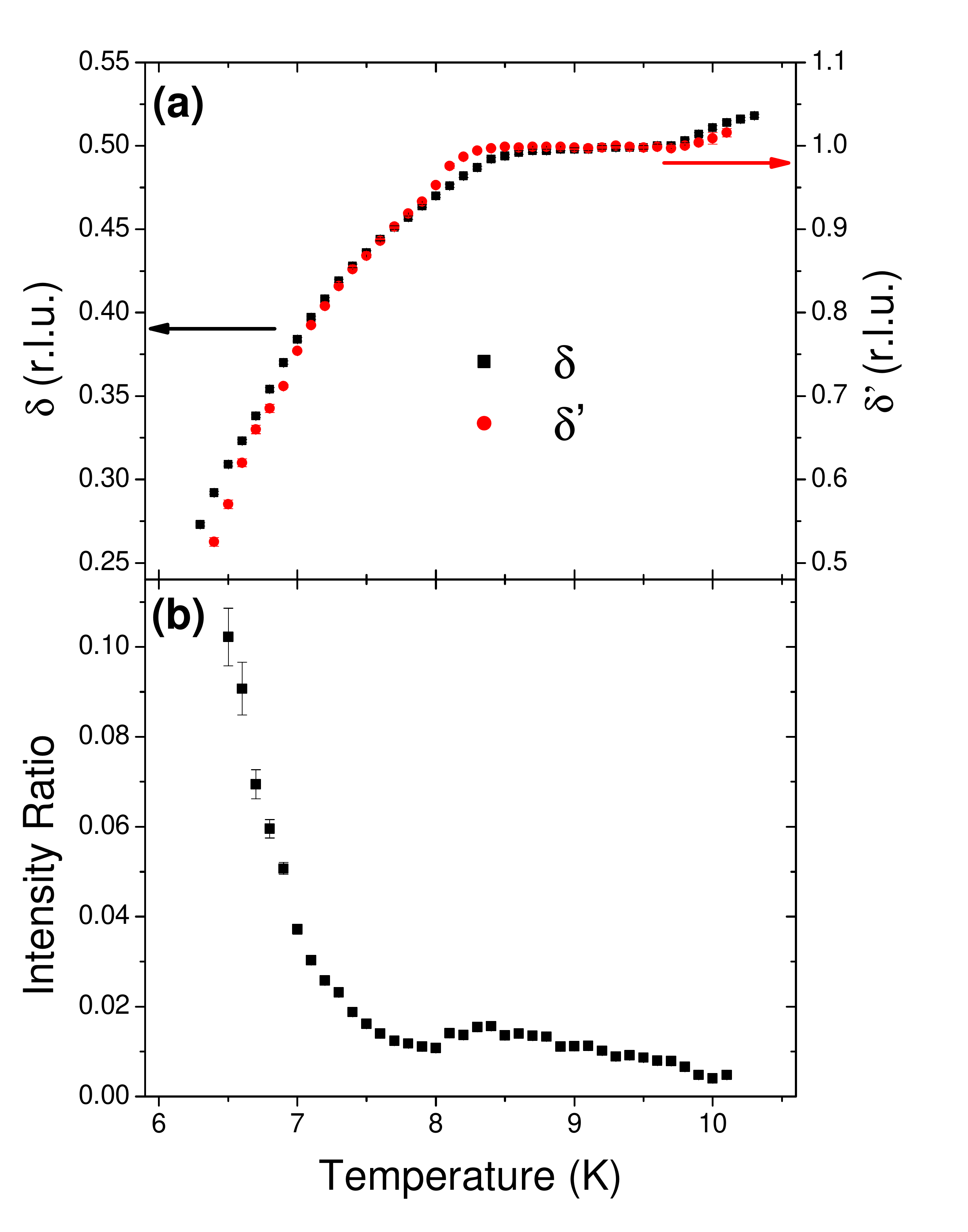} \vspace{-4mm}
\caption{(a) Positions for the primary and second-order peaks, shown as $\delta$ (left scale) and $\delta^{\prime}$ (right scale), measured at $\mu_{0}H$~=~0.10~T. The scale for the second-order peak is double that of the first-order peak, showing that $\delta^{\prime}$~$\approx$~2$\delta$. (b) Ratio of the second-order and first-order peak integrated intensities, also measured at $\mu_{0}H$~=~0.10 T.  The second-order relative intensity decreases monotonically with increasing temperature, except for an increase upon entering the $\delta^{\prime}$~=~1.0 commensurate lock-in phase.  Uncertainties throughout this article are statistical and refer to one standard deviation.} \vspace{-2mm}
\label{HigherOrder}
\end{figure}

In Figure~\ref{HigherOrder}(a), the positions of the primary and higher-order peaks are shown for the data measured at $\mu_{0}H$~=~0.10~T.  The left scale denotes the $\delta$ value for the first-order peak, while the right scale denotes the $\delta^{\prime}$ value for the higher-order peak; the scale for the higher-order peak is double that of the primary peak, so that the overlap of these data demonstrates that $\delta^{\prime}$~$\approx$~2$\delta$.  This matches the results from the $\mu_{0}H$~=~0.30~T data where $\delta^{\prime}$ values of 0.8 and 1 were observed in the $\delta$~=~2/5 and 1/2 phases, respectively, demonstrating that all higher-order reflections in a magnetic field are second-harmonics.  Figure~\ref{HigherOrder}(b) displays the ratio of the second-order and first-order integrated intensities in the $\mu_{0}H$~=~0.10~T data.  The relative intensity of the second-order reflection generally decreases with increasing temperature, demonstrating that the magnetic SDW is closer to a perfect sinusoid at higher temperatures with the squaring-up becoming more pronounced at lower temperatures; however, the second-order intensity is also slightly enhanced in the temperature range corresponding to the $\delta$~=~1/2 phase.

\begin{figure}
\centering
\vspace{2mm}
\includegraphics[width=7.6cm]{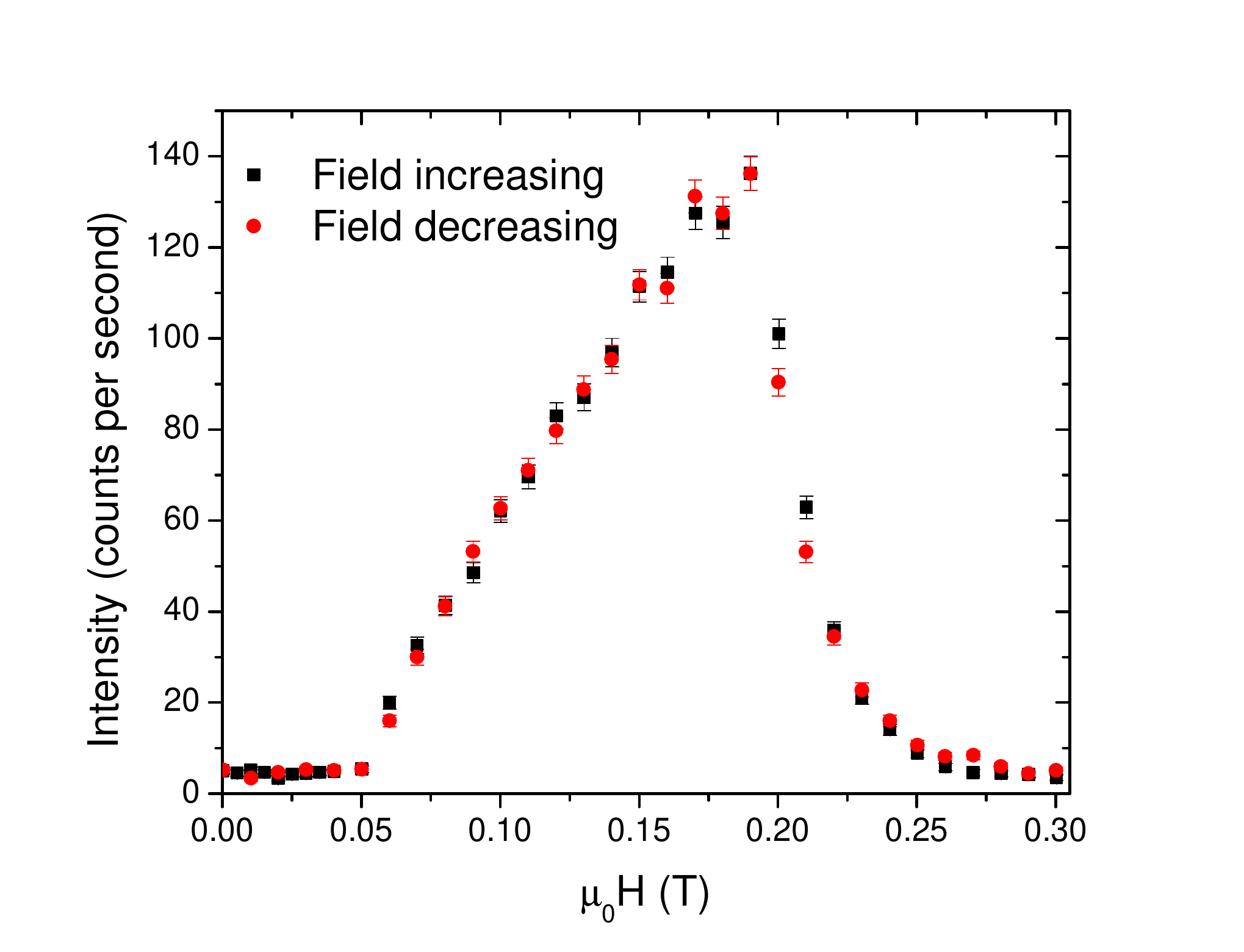} \vspace{-4mm}
\caption{Field dependent intensity of the (0~3~0) reflection measured at $T$~=~7.9~K with increasing and decreasing field.  This second-order reflection arises sharply at 0.05~T.} \vspace{-2mm}
\label{HO3}
\end{figure}

Figure~\ref{HO3} displays the field dependence of the intensity of the (0~3~0) reflection, which corresponds to the second-order scattering in the $\delta$~=~1/2 phase, measured at 7.9~K on the small single crystal.  This temperature lies roughly in the middle of the narrow $\delta$~=~1/2 phase measured at a constant field of 0.05~T.  The intensity of this reflection appears sharply at a minimum field of 0.05~T; importantly, there is no measurable magnetic scattering at the (0~3~0) position for any field smaller than 0.05~T despite the fact that the material is also in a $\delta$~=~1/2 phase at this temperature in zero field.  This suggests that the two regions of the phase diagram displaying a $\delta$~=~1/2 phase should be considered distinct from one another, as only the high-field region displays second-order scattering.  The intensity of this reflection falls off at higher fields, signifying a transition into the $\delta$~=~2/5 phase at a field of about 0.2~T.  The field dependence of the intensity at the (0~2.5~0) reflection at 7.9~K showed a narrow dip in intensity at 0.05~T, but at this temperature any discommensuration comparable to that reported at 7.5~K\cite{Petrenko} was too small to be clearly resolved.

\section{SUMMARY}

In summary, Co$_{3}$V$_{2}$O$_{8}$ features a variety of magnetic phases arising from competing magnetic interactions on a frustrated lattice.  The application of a magnetic field along the magnetization easy-axis promotes the development of a net moment; this adds yet another term in an already complicated Hamiltonian.  Just as competing interactions in zero field yield multiple different phases as a function of temperature, the impetus to produce a net magnetization under field yields a complex phase diagram.  Small applied fields, $\mu_{0}H$~$\approx$~0.05~T, destabilize the low-field commensurate structures in favor of the incommensurate spin density wave.  Larger applied fields lead to the restoration of a $\delta$~=~1/2 commensurate phase, a new commensurate SDW with $\delta$~=~2/5, and second-harmonic higher-order scattering.  Scattering at the (0~3~0) reflection arises sharply at a field of 0.05~T, suggesting that the low-field and high-field $\delta$~=~1/2 regions of the phase diagram differ in that only the latter displays a squaring-up of the sinusoidal SDW and the concomitant second-order reflections.

\section*{ACKNOWLEDGEMENTS}

We thank Brooks Harris, Taner Yildirim, and Amnon Aharony for helpful discussions.  J.S.H. acknowledges support from the NRC/NIST Postdoctoral Associateship Program. The work in Minsk was partly supported by the Belarusian Fund for Basic Scientific Research, grant number F10R-154.\\
\\
* email: joel.helton@nist.gov\\
$\ddag$~email: jeffrey.lynn@nist.gov\\
$\dag$~Present address: DuPont Central Research and Development, Experimental Station, Wilmington, DE 19880
\bibliography{CVO}
\end{document}